\pdfoutput=1 %
\documentclass{lipics-v2021}
\usepackage{latexsym}
\usepackage{amsmath}

\usepackage{microtype}

\title{Natural Language Specifications in Proof Assistants} %
\hideLIPIcs

\author{Colin S. Gordon}{Drexel University, USA \and \url{https://www.cs.drexel.edu/~csgordon/}
}{csgordon@drexel.edu}{https://orcid.org/0000-0002-9012-4490}{}%

\author{Sergey Matskevich}{Drexel University, USA}{sm3372@drexel.edu}{}{}

\authorrunning{C.~S.~Gordon and S.~Matskevich}%

\Copyright{Colin S. Gordon and Sergey Matskevich}%

\nolinenumbers %

\usepackage{mathpartir}
\usepackage{graphicx}
\usepackage{xcolor}
\usepackage{listings}
\lstdefinelanguage{scala}{
  morekeywords={abstract,case,catch,class,def,%
    do,else,extends,false,final,finally,%
    for,if,implicit,import,match,mixin,%
    new,null,object,override,package,%
    private,protected,requires,return,sealed,%
    super,this,throw,trait,true,try,%
    type,val,var,while,with,yield},
  otherkeywords={=>,<-,<\%,<:,>:,\#,@},
  sensitive=true,
  morecomment=[l]{//},
  morecomment=[n]{/*}{*/},
  morestring=[b]",
  morestring=[b]',
  morestring=[b]"""
}

\lstdefinelanguage{Coq}{
    morekeywords={Proof,Qed,Polymorphic,Lemma,Hint,Unfold,Local,Obligation,Tactic,Ltac,Instance,Class,Defined,Require,Import,Section,End,Parameter,Record,Program,Example,Inductive,Infix,Notation,Definition,Fixpoint,Postulate,Goal},
    morekeywords=[2]{eauto},
    keywordstyle=[2]\color{teal},
    morekeywords=[3]{Set,Prop,Type,with,forall,using,fun},
    keywordstyle=[3]\color{olive},
    otherkeywords={->,|,:,=},
    sensitive=true,
    morecomment=[n]{(*}{*)},
    basicstyle=\small\ttfamily,
    keywordstyle=\color{blue},
    commentstyle=\color{violet}\small\ttfamily}
\usepackage{xspace}
\usepackage{stmaryrd}
\usepackage{amsmath}

\newcommand{\coq}{\textsc{Coq}\xspace}

\begin{document}

\maketitle
\begin{abstract}
  Interactive proof assistants are computer programs carefully constructed to check a human-designed proof of a mathematical claim with high confidence in the implementation.
  However, this only validates truth of a formal claim, which may have been mistranslated from a claim made in natural language.  This is especially problematic when using proof assistants to formally verify the correctness of software with respect to a natural language specification.
  The translation from informal to formal remains a challenging, time-consuming process that is difficult to audit for correctness.
  This paper argues that it is possible to build support for natural language specifications within existing proof assistants, in a way that complements the principles used to establish trust and auditability in proof assistants themselves.
\end{abstract}

\section{Introduction}
Proof assistants can establish very high confidence in the correctness of formal proofs, both because of their rigorous checking and attention to producing independently auditable evidence that the arument is correct~\cite{paulson1990logic,pollack1998believe}.
But one of the unavoidable points of trust for even a carefully-implemented proof assistant is the specifications themselves: proving the wrong theorem is of limited use. And only those who can read both formal and informal specifications can even consider whether this has occurred. This is particularly crucial for software verification: 
software specifications typically originate in natural language, and any accompanying formal specification comes afterwards --- which increasingly occurs for compilers~\cite{leroy2009formally}, operating systems~\cite{seL4TOCS}, and other high-value software.
Currently, the \emph{only} bridge between the formal and informal specifications is the humans who perform the translation. 
There is no independently checkable record of this translation aside from the possibility of comments or notes by the translators --- themselves largely in informal (though likely careful) natural language.
And simply being familiar with both the specification language and the intended specification is insufficient by itself to bridge this gap~\cite{finney1996empirical,finney1996mathematical}: relating the two is a separate skill that is independently challenging to develop.

Ideally, it would be possible to give natural language specifications directly to the proof assistant, for example:
\begin{lstlisting}[language=Coq]
Goal "addone is monotone".
\end{lstlisting}
Robust support for such specifications could enable significant improvements in requirements tracing (machine \emph{checked} mappings from natural language to formal results), including for artifact evaluation; education, where it could help students check their understanding of how either mathematics or program specifications are formalized logically; and even communication with non-technical clients who might wish to have some confidence that a formalization they do not themselves fully understand is correct. On the last point, Wing's classic paper introducing formal methods~\cite{wing1990specifier} posits that customers may read the formal specifications produced from informal requirements, but this is only possible if the client can make sense of the formal specification itself. Machine-checked relationships between natural-language and formalized expressions of software properties can help bridge this knowledge gap by connecting formal properties to natural language a reader with less background in a specific formal logic could understand.

By the end of this paper, we will have developed a way to accept the very similar
\begin{lstlisting}[language=Coq]
Goal spec "addone is monotone".
\end{lstlisting}
(where \lstinline|spec| returns the logical form of the natural language utterance in quotes).
We envision such a system can eventually be used for the purposes above, to generate formal claims about mathematics or programs verified in a proof assistant, whether specified in a proof assistant's own logic, or indirectly through a foundational program logic~\cite{appel2001foundational} built inside a proof assistant. While these generated specifications may not match the expert human formalizer's choices (often influenced by setting up definitions in a particular way to simplify proofs), they should be implied by the specification used to conduct the primary verification activities, offering an additional, optional level of assurance.

This is not a job for machine-learning-centric natural language processing, which is incompatible with the goals of using a proof assistant for formal verification. inappropriate for foundational verification. There is no guarantee a learned translation is sensible, and if a translation from natural to formal language ends up being surprising, few machine learning approaches produce an auditable trail of evidence for \emph{why} that translation is believed correct (in the eyes of the trained model), and no way to precisely fix misunderstandings of specific words. Meanwhile, \emph{proof certificates} play a central role in the design of trustworthy proof assistants~\cite{pollack1998believe,gordon2000lcf} and foundational program verification~\cite{appel2001foundational}.
Moreover, as proof assistants are often used to formalize properties of new mathematics or new programs, often using new terminology, there will often be a lack of training data for mapping natural language to a formal property.
Later, we point out additional ways that the needs of trusted formalization of natural language specifications run afoul of many of machine learning's known limitations, while requiring few of its advantages. (We also point out limited ways machine learning can play a role in optimizing the techniques we employ.)

Fortunately the field of linguistics predates machine learning. Formalizing \emph{categorial grammar}~\cite{ajdukiewicz1935syntaktische,bar1953quasi,lambek1958mathematics,Steedman:2001} carefully in a proof assistant offers a path to a natural, auditable way to bridge the gap between formal and natural-language specification.
In this paper we show a prototype demonstrating that it is possible parse a string containing a natural language specification into a semantic representation that can be used directly in proofs within a proof assistant (i.e., a proposition in \coq's logic) in a principled way using \coq's typeclasses~\cite{sozeau2008first}, and argue that this approach is modular and can extend to sophisticated natural language.
A partial adaptation to the Lean theorem prover suggests that with Lean's algorithmic improvements~\cite{selsam2020tabled} this can be made efficient enough for interactive use.
We also analyze how the trusted computing base is affected when considering trust of a formal verification up to the natural language specification.

This paper establishes that the theoretical core is readily within reach. But the robust realization of these ideas posited above will require significant efforts to develop a broad core vocabulary to act as a starting point for extensions (we outline how such extensions would work) and eventually a robust library of domain-specific language support. It will require time and collective effort to collect precise natural language descriptions of partial specifications to guide development and evaluation of such a system. And it raises potential opportunities for fruitful collaboration with linguists on building systems that simultaneously benefit consumers of proof assistants while offering a live proving ground for work on semantic parsing.

  \section{Background \& Motivation}\label{background}

  This section provides a condensed (and therefore somewhat biased)
  background in natural language processing and categorial grammar, from a programming languages point of view.
  
  \emph{Categorial grammar} is a body of work concerned with the use of
  techniques and ideas from logic to relate the syntax of natural language
  with the meaning of language in a way independent of syntax.  The core idea is to build a sort of type theory where base types correspond to grammatical \emph{categories} (hence \emph{categorial}), from which more complex grammatical categories can be defined.  A set of inference rules is then used to define, simultaneously, how grammatical categories combine into larger sentence fragments and how those smaller fragments' meanings (logical forms) are combined into larger meanings. This process bottoms-out at a \emph{lexicon}, giving for each word its grammatical roles (types) and associated denotations. Thus categorial grammar is a system of simulatenously parsing natural language from strings and assigning denotational semantics --- a process traditionally referred to as \emph{semantic parsing} in the computational linguistics literature.

  Most prominent in the linguistics community are \emph{combinatory categorial grammars} (\textsc{CCG}s)~\cite{Steedman:2001}, though also relevant to our goals are the \emph{categorial type logics} (\textsc{CTL}s\footnote{Occasionally also called \emph{type-logical grammars} (\textsc{TLG}s).}).  Work on \textsc{CCG}s epmhasizes appropriate constructs for linguistic ends, while \textsc{CTL}s hew close to Lambek's view~\cite{lambek1958mathematics} of categorial grammars as substructural logics for linguistics. 
  While these reflect very different philosophical and practical aims, for our present purposes the distinction is immaterial: it is widely held, and in some cases formalized~\cite{kruijff2000relating,gerhard2005anaphora}, that rules used in \textsc{CCG}s (including the variant with the most sophisticated linguistic treatments~\cite{Baldridge:2003:MCC:1067807.1067836}) correspond to theorems in particular \textsc{CTL}s~\cite{Moortgat1996}.
  In this work we use only principles common to both \textsc{CCG}s and \textsc{CTL}s.

  All categorial grammars parse by combining sentence fragments based on their grammatical types.  These types include both atomic primitives (such as noun phrases) as well as more complex types, namely so-called \emph{slash-types} that indicate a predicate argument structure (which are used to model, for example, most classes of verbs).
  Oversimplifying slightly, categorial grammars treat parsing as logical deduction
   in a residuated non-commutative linear logic.\footnote{Technically only \textsc{CTL}s~\cite{morrill2012type} take this as an epistemilogical commitment, while \textsc{CCG}s~\cite{Steedman:2001}) are agnostic, inheriting such a relation via Baldridge and Kruijff's work~\cite{Baldridge:2003:MCC:1067807.1067836}.} This is essentially a family of linear logics without the structural rule for freely commuting the order of assumptions, thus modeling sensitivity to word order, and picking up as a consequence \emph{two} forms of implication corresponding to whether an implication expects its argument to the left or to the right.\footnote{It is the presence of the ability to commute assumptions arbitrarily that allows a single implication to suffice in standard logics.}
  The model for the logic is a sequence of words, and types correspond to
  the grammatical role of a sentence fragment. 

  $A/B$ ($A$ over $B$) is the grammatical type for a fragment that, when given a $B$ to its right, forms an $A$.  $A\backslash B$ ($A$ under $B$) is the grammatical type for a fragment that, when given an $A$ to its left, forms a $B$.  In both cases, the argument is ``under'' the slash, and the result is ``above'' it.\footnote{We follow CTL notation rather than CCG notation (which always puts results to the left) as users of proof assistants tend to be familiar with a range of logics, so the CCG syntax would likely confuse users already familiar with the Lambek calculus and related systems. This notational choice is orthogonal to the choice of which rules to employ.})
  These are called \emph{slash types}.
  The grammars include rules to combine adjacent parts of a sentence.
  The elimination rules are the first two in Figure \ref{fig:rules}.
  \begin{figure*}
  \begin{mathpar}
    \inferrule[$\backslash$-Elim]{\Gamma\vdash A \Rightarrow a\quad \Delta\vdash A\backslash B \Rightarrow f}{\Gamma,\Delta\vdash B \Rightarrow (f~a)}
    \and
    \inferrule[/-Elim]{\Gamma\vdash A/B \Rightarrow f\quad \Delta\vdash B \Rightarrow a }{\Gamma,\Delta\vdash A \Rightarrow (f~a)}
    \and
  \inferrule[/-Comp]{\Gamma\vdash A/B \Rightarrow e \quad \Delta \vdash B/C \Rightarrow f }{\Gamma,\Delta\vdash A/C \Rightarrow (e\circ f)}
  \and
  \inferrule[$\backslash$-Comp]{\Gamma\vdash A\backslash B \Rightarrow e \quad \Delta \vdash B\backslash C \Rightarrow f }{\Gamma,\Delta\vdash A\backslash C \Rightarrow (f\circ e)}
  \and
  \inferrule[Reassoc]{
    (\Gamma,\Delta),\Upsilon\vdash A \Rightarrow e
  }{
    \Gamma,(\Delta,\Upsilon)\vdash A \Rightarrow e
  }
  \and
  \inferrule[Shift]{
    \Gamma\vdash A \backslash (B / C) \Rightarrow f
  }{
    \Gamma\vdash (A \backslash B) / C \Rightarrow (\lambda r,l\ldotp f~l~r)
  }
  \end{mathpar}
  \vspace{-1.5em}
  \caption{A selection of rules used in this paper, all derivable in \textsc{CTL}s and \textsc{CCG}s.
  }
  \label{fig:rules}
  \end{figure*}
  The judgment $\Gamma\vdash C \Rightarrow e$ is read as claiming the sequence of words $\Gamma$ can be combined to form a sentence fragment of grammatical type $C$, whose underlying semantic form --- logical form --- is given by $e$.  $e$ is a term drawn from the logical language being used to represent sentence meaning, typically a simply-typed lambda calculus in keeping with Montague~\cite{montague1970universal,montagueEFL70,montaguePTQ73,partee1997montague}, though in our work we follow the alternative approach~\cite{bekki14dts,sundholm1986proof,chatzikyriakidis2017modern,ranta1994ttg} of targeting a dependently-typed calculus.

  A \emph{lexicon} gives the grammatical role and semantics for individual words, providing the starting point for combining fragments.
  Categorial grammars push all knowledge specific to a particular human language into the lexicon, in categorizing how individual words are used. This allows the core principles to be reused across languages, which has been put to use in building wide-coverage lexicons for a variety of natural languages~\cite{hockenmaier2007ccgbank,hockenmaier2006creating,ambati2018hindi,abzianidze2017parallel,moot2015type}.
  
  Together, these allow filling in choices for the metavariables in the rules above, which together permit derivations like
\[
\inferrule*[right=$\backslash$-Elim]{
    \inferrule*{ }{\textrm{``four''}\vdash NP \Rightarrow 4}\;
    \inferrule*[right=/-Elim]{
      \inferrule{ }{\textrm{``is''} \vdash (NP\backslash S)/ADJ \Rightarrow \lambda p\ldotp\lambda x\ldotp p~x}\;
      \inferrule{ }{\textrm{``even''} \vdash ADJ \Rightarrow \mathsf{even}}
    }{\textrm{``is~even''}\vdash NP\backslash S\Rightarrow (\lambda x\ldotp \mathsf{even}~x)}
}{
    \textrm{``four~is~even''}\vdash S\Rightarrow \mathsf{even}~4
}
\]
  Each grammatical type $C$ corresponds to a particular type in the underlying lambda calculus and the underlying \emph{semantic} type is determined by a systematic translation from the syntactic grammatical type.
  Borrowing more notation from logic (where this idea is known as a Tarski-style universe~\cite{martin1984intuitionistic}), we write $\lfloor C\rfloor$ for $C$'s semantic type.  
  Both slash types correspond to function types in the lambda calculus: 
  $ \lfloor A/B\rfloor = \lfloor B\backslash A\rfloor = B\rightarrow A$.
  An invariant of the judgment $\Gamma\vdash C\Rightarrow e$ is that in the underlying logic, $e$ always has type $\lfloor C\rfloor$.
  This invariant explains why it is correct for ``four'' to have semantics $4$ while ``is even'' has a function as its semantics.
    In proof assistants based on type theory, like \coq, the set of grammatical types can be given as a datatype declaration, and the interpretation function as a function from grammatical types to proof assistant types.

  Figure \ref{fig:rules} includes a selection of additional rules,
  each of which is either an axiom or derived rule in CCGs, and a theorem in the Lambek calculus.  Thus we do not commit to one approach over the other at this time.

  These few rules are enough to formalize a small fragment of English, and demonstrate the possibility of interpreting natural language specifications \emph{within} a proof assistant, in a well-founded and extensible way.  Our initial choice of rules is limited, but not fundamentally so. \textsc{CCG}s recognize \emph{mildly context-sensitive languages}~\cite{joshi:91,Kuhlmann2015}, which are believed to cover the full range of grammatical constructions in any natural language~\cite{steedman2012taking}. All rules of \textsc{CCG}s are encodable in the way we describe, as are the rules of Turing-complete \textsc{CTL}s~\cite{carpenter1999turing}. Ultimately the question of which rules are required is an empirical one, considering both linguistic constructions and the complexity of recognizing these grammars using typeclass search.
  For now we are adding additional \textsc{CCG} rules as needed.

  \section{Exploring Categorial Grammars for \coq Specifications}\label{open-categorial-grammars-over-cic}
  This section describes a model of a very small fragment of English for describing simple mathematics.
  Our goal is not to present a polished and complete natural language fragment suitable for a wide range of specifications.  Producing such a result is a very long-term goal, elaborated in Section \ref{sec:treebanks}.
  The purpose of this section then is to emphasize:
  \begin{enumerate}
    \item that existing work on linguistic semantics covers many of the grammatical aspects of the language constructions we use when speaking or writing about formal claims, and is sufficiently flexible to admit extensions for grammar unique to mathematical prose; 
    \item that existing proof assistants (by example \coq, but by association similar systems like \textsc{Lean}) offer an environment ready to implement linguistic semantics in a way directly integrated into the use of specifications in proof assistants; and
    \item extensions of this idea are well worth investigating.
  \end{enumerate}

  We propose using \coq's typeclass support~\cite{sozeau2008first} to perform semantic parsing from natural language\footnote{In this paper, English, but in principle any other natural language with thorough treatments in categorial grammar~\cite{hockenmaier2007ccgbank,hockenmaier2005ccgbank,hockenmaier2006creating,abzianidze2017parallel,ambati2018hindi,moot2015type,mineshima2016building}.} to \coq's specification logic, or alternatively (see Section \ref{sec:generalization}) embedded logics.
  This approach naturally supports an open-ended lexicon, which is essential to modularly extending the words handled by semantic parsing (Section \ref{sec:treebanks}).  Moreover, it has the advantage over external tools that it works within existing proof assistants today, with no need to involve an external toolchain for the translation. This ensures the extended lexicon can grow in tandem with a formal development, with organization chosen by developers rather than dictated by an external tool, and with \coq automatically checking validity of the lexicon extensions at the same time it checks validity of the rest of the formalization.

  Most work on categorial grammars lumps all entities --- frogs, people, books --- into a single semantic type.  But for logical forms to talk about entities in \coq's logic --- which distinguishes natural numbers, rings, monoids, and so on with different types --- adjustments must be made.
  Noun phrases and related syntactic categories must be parameterized by the semantic (\coq) type of entity they concern.
  Adapting categorial grammar to refer to objects in \coq's logic --- a dependently-typed lambda calculus known as the Calculus of Inductive Constructions (\textsc{CIC})~\cite{paulin1993inductive} requires making the base grammatical categories multi-sorted, distinguishing nouns with different semantic types.
    Traditional categorial grammars assume the logic used for logical forms has a sort $E$ for ``entities'', which different kinds distinguished by predicates: e.g., $\textsf{three} : E$ and $\textsf{Mary}:E$, but $\mathsf{number}(\textsf{Mary})=\textsf{false}$.  This approach is rooted in the assumption that first order logic is an appropriate semantic model for sentence meaning.
    That assumption is plausible for many general circumstances, and useful for modeling figurative speech, but is clearly incompatible with making natural language claims about mathematical objects defined in intuitionistic type theory.

    We follow linguistically-motivated work using intuitionistic type theories like \coq's for exploring possible logical forms~\cite{sundholm1986proof,ranta1991intuitionistic,ranta1994ttg,chatz2014nlcoq,bekki14dts,bekki2012LIRA,ranta1995context}, in using an alternative model where many grammatical categories are indexed by the underlying semantic type to which they refer. 

  \coq's logic is expressive enough to give the set of grammatical types as a datatype \lstinline|Cat|, and to give the interpretation of those types into semantic types as a recursive function \lstinline|interp| within \coq, as shown in Figure \ref{fig:corecoq}.

\begin{figure}
\begin{lstlisting}[language=Coq]
Inductive Cat : Type :=
  | S (* Sentence/proposition *)
  | NP : forall {x:Type}, Cat
  | rSlash : Cat -> Cat -> Cat (* A/B *)
  | lSlash : Cat -> Cat -> Cat (* A\B *)
  | ADJ : forall {x:Type}, Cat
  | CN : forall {A:Type}, Cat.

Fixpoint interp (c : Cat) : Type :=
match c with 
| S => Prop (* Coq's type of propositions *)
| @NP t => t
| rSlash a b => interp b -> interp a
| lSlash a b => interp a -> interp b
| @ADJ t => t -> Prop
| @CN t => unit
end.
\end{lstlisting}  
\caption{Core grammatical categories as a \coq type \textsf{Cat} and the mapping $\lfloor-\rfloor$ from grammatical types to semantic types as \textsf{interp}.}
\label{fig:corecoq}
\end{figure}

Grammatical types (categories) \textsf{Cat} (categories) include the aforementioned slash types, sentences ($S$); 
noun phrases ($NP~A$) denoting objects of \coq type $A$; adjectives ($ADJ~A$) denoting predicates over such objects; and common nouns ($CN$) denoting unit types, but imposing constraints on the semantic type indices of other phrases in the context of a sentence, used in cases where a sentence must refer to a common class of objects (i.e., a type), such as ``natural numbers'' or ``rings.'' . 
As mentioned previously, the slash types correspond to function types (the direction is relevant only in the grammar, not the semantics), sentences are modeled by \coq propositions (the type of logical claims), noun phrases of \coq type \texttt{t} correspond to elements of \texttt{t}, and similarly adjectives correspond to predicates on such types (i.e., elements of \lstinline[language=Coq]|t -> Prop|).
These are modeled by the \coq function \lstinline|interp|, which maps grammatical types to other \coq types.
Common nouns denote as the unit type. This may seem odd, but common nouns essentially serve as a means to force certain type variables to be the ones corresponding to a specific word.\footnote{In traditional semantics in first-order logic, common nouns often denote a kind of predicate to guard quantifications.}

We use \coq's notation facilities (essentially, macros) to write the slash types as \lstinline|A // B| for $A / B$, and \lstinline|A \\ B| for $A \backslash B$.
We also use these macros to define more interesting grammatical categories.\footnote{We could use \coq definitions as well, but macros work better with the unification in the next section.}
For example, a quantifier over a certain common noun type is given as:
$ \mathsf{Quant}~A \equiv  (S / (NP_A \backslash S) / CN_A) $.%
That is, a quantifier discussing a \coq type $A$ looks first to its right for a common noun (which constrains the type $A$ to the one named by the natural language common noun). Then the result of that looks to its right for a sentence fragment expecting a noun phrase to its left.  The latter sentence fragment is essentially a predicate.

  \label{a-categorial-grammar-dsl-in-coq}
\subsection{Combination Rules}
Parsing natural language specifications requires automatically applying rules like \textsc{/-Elim} to combine sentence fragments.
Rather than modifying \coq, we can use existing trusted\footnote{Officially, typeclasses are not part of \coq's trusted computing base, as they elaborate to record operations before being passed to the core proof checking apparatus. In practice, they mediate \emph{which} terms are passed to the core, so calling them \emph{untrusted} would be a misnomer.} functionality to do this for us: typeclasses~\cite{sozeau2008first}. These are a mechanism for parameterizing function definitions by a set of (often derivable) operations.  \coq permits declaring a typeclass (roughly, an interface), and declaring implementations associated with certain types.  The implementations may be parameterized by implementations for other types (such as defining an ordering on pairs in terms of orderings for each component of the pair).
  When a function is called that relies on a set of operations, \coq attempts to use higher-order unification to construct an appropriate implementation.
  It is possible to encode the rules of a system like a CTL or CCG into typeclasses.
  Each judgment signature corresponds to a typeclass, and each rule corresponds to an instance (implementation) of the typeclass.
  We define the judgement form \(\Gamma\vdash T \Rightarrow e\) as:
  
\begin{lstlisting}[language=Coq]
Class Synth (l : list word) (cat : Cat) := 
  { denote : interp cat }.
\end{lstlisting}
If an instance of \lstinline|Synth l C| exists, it comes with an operation \lstinline|denote| that produces a \coq value of type \lstinline|interp C| --- $\lfloor C\rfloor$.
 Because \(e\) is viewed as an output we would like
  to query, it is defined as a member of the typeclass, rather than as an
  additional index. When deriving a formal specification for sentence
  \(s\), we will arrange for the typeclass machinery to locate an instance
  of \texttt{Synth\ s\ S} --- checking that \(s\) is a grammatically valid
  sentence --- and request its term denotation when necessary.
 Translating a specification given by the list of words \lstinline|w| corresponds to parsing \lstinline|w| as a sentence: finding an instance of \lstinline|Synth w S|.
 We define an instance for each rule to encode, 
  such as this one 
  corresponding to \textsc{\textbackslash-Elim}, which applies the logical form of the functional to the logical form of the argument:
\begin{lstlisting}[language=Coq,mathescape]
Instance SynthLApp {$\Gamma$ $\Delta$ A B}
 (L:Synth $\Gamma$ A)(R:Synth $\Delta$ (A \\ B)) :
 Synth ($\Gamma$ ++ $\Delta$) c2 := 
{denote := (denote R) (denote L)}.
\end{lstlisting}
and for leftward composition (\textsc{\textbackslash-Comp}):
\begin{lstlisting}[language=Coq,mathescape]
Instance LComp:{$\Gamma$ $\Delta$ A B C}
 (L:Synth $\Gamma$ (A\\B))(R:Synth $\Delta$ (B\\C))
  : Synth ($\Gamma$ ++ $\Delta$) (A \\ C) := 
{denote := fun x1 => (denote R) (denote L x1)}.
\end{lstlisting}
The remaining rules of Figure \ref{fig:rules} can also be encoded in this way.

In addition to exploiting the proof assistant's built-in search for parsing, the use of typeclasses means the set of rules is extensible. As mentioned, the core \textsc{CCG} rules are already known to be expressive enough to cover any known linguistic construction (the technical term is that \textsc{CCG}s are \emph{mildly context-sensitive}~\cite{joshi:91,Kuhlmann2015}). However:
\begin{itemize}
  \item Additional derived rules could be added either to accelerate proof search for recurring intricate constructions
  \item The rules can be used to offer generalized constructions for words with many roles. This is particularly valuable for linguistic constructs likc coordination (Section \ref{sec:coord}).
  \item As discussed momentarily, it allows easy extension of the lexicon with new words without modifying a fixed database.
\end{itemize}

\subsection{Lexicon}
The lexicon is encoded via another typeclass which assigns grammatical types and logical forms to individual words rather than series of words.
\coq permits declaring multiple instances for the same word (e.g., if a word has multiple meanings of different grammatical types), giving essentially a free variant of intersection types~\cite{moortgat1999constants} without the coherence issues described by Carpenter~\cite{carpenter1997type} (only one definition will be chosen per appearance of the word).
  We represent our lexicon with another type class, and tie it into the \lstinline|Synth| typeclass:
\begin{lstlisting}
Class lexicon (w : word) (cat : Cat) := { denotation : interp cat }.
Instance SynthLex {w cat}`(lexicon w cat) : Synth [w] cat := 
  { denote := denotation }.
\end{lstlisting}

  Thus, a dictionary for our approach consists of a set of instance declarations for \lstinline|lexicon|:
\begin{lstlisting}
Instance fourlex : lexicon "four" NP := { denotation := 4 }.
Instance noun_is_adj_sentence {A:Type} : 
  lexicon "is" (@NP A \\ (S // @ADJ A)) :=
   { denotation n p := p n }.
Instance noun_is_noun_sentence {A:Type} : 
  lexicon "is" (@NP A \\ (S // @NP A)) := 
   { denotation n a := n = a }.
\end{lstlisting}
Here we have defined two different meanings for ``is'' allowing it to be used to apply an adjective, or to denote equality.  The difference between the two, beyond their denotation is the grammatical types: both expect a noun phrase to the left, and some other word to the right: an adjective in the first case, or another noun in the second.
Note that in both cases, the adjective or noun phrase must match the type of underlying \coq object the the left-side noun phrase refers to: the argument \lstinline|n| is in both cases a variable of type \lstinline|interp (@NP A)=A| because that is the argument of the outermost slash type, while the second argument in each entry corresponds to the interpretation of the second slash type's argument (a predicate or an additional term, respectively).
Coupled with a development-specific bit of lexicon to name a particular \coq object of interest:
\begin{lstlisting}[mathescape]
Instance addone_lex : lexicon "addone" NP :=
  { denotation := addone (* $\lambda$x. x + 1 *) }.
\end{lstlisting}
this approach permits giving correct denotations to both:
\[\begin{array}{c}\llbracket\textrm{addone is monotone}\rrbracket \equiv \mathsf{monotone}(\texttt{addone})
\\
\llbracket\textrm{addone given 3 is 4}\rrbracket \equiv (\texttt{addone}\;3)=4
\end{array}\]

\subsubsection{Quantifiers}
\label{sec:quantifiers}
Earlier we mentioned quantifiers over $A$ can be given grammatical type 
\[\mathsf{Quant}~A \equiv  (S / (NP_A \backslash S) / CN_A) \]
Thus, a quantifier looks to its right first for a common noun (corresponding to the word identifying the \coq type to quantify over), and after that is combined, the result looks further to the right for a sentence fragment expecting such a thing to its left.
Then adding appropriate lexicon entries for ``every'':
\begin{lstlisting}
Instance forall_lex {A:Type} : lexicon "every" (Quant A) :=
{ denotation := fun _ P => (forall (x:A), P x) }.
\end{lstlisting}
and another for the common noun ``natural'' (number) allows correctly parsing sentences like
\[\llbracket\textrm{every natural is even}\rrbracket\equiv\forall (n:nat)\ldotp (\mathsf{even}\;n)\]
(Recall, we must still be able to state claims that are false.)
The common noun contributes nothing directly to the denotation, but constrains the quantifier to work with noun phrases referring to natural numbers.

\subsubsection{Coordination}
\label{sec:coord}
One aspect of natural language which is the source of some interest is that the words ``and'' and ``or'' (or their equivalents in other languages) can often be used to combine sentences fragments of widely varying grammatical types.
For example, in ``four is even and positive'' the word ``and'' conjoins two adjectives: ``even'' and ``positive.''
Yet in the sentence ``four is even and is positive'' it conjoins two phrases of grammatical type $NP_\textsf{nat}\textrm{\textbackslash} S$ (``is even'' and ``is positive'').

We can directly adopt a solution from the compuational linguistics literature~\cite{carpenter1997type}, and formalize that ``and'' and ``or'' apply to any semantic type that is a function into (a function into\ldots) the type \lstinline|Prop| of propositions.  
We define an additional typeclass to recognize such ``Prop-like'' grammatical types inductively, starting with the grammatical types $S$ and $ADJ$, and inductively including slash types whose result type is also ``Prop-like'', which define an operation to lift boolean semantics through repeated functions.
We then add a polymorphic lexicon entry for each of ``and'' and ``or'' which assigns them any ``Prop-like'' type.

Thus in a sentence like ``four is even and is positive'' the two conjuncts are recognized as Prop-like (their underlying semantic type is $\textsf{nat}\rightarrow\mathsf{Prop}$), and the operations of the typeclass recognizing this automatically lift a binary operation on \lstinline|Prop| to a binary operation on predicates. For ``and'' this lifts logical conjunction to $\lambda P\ldotp\lambda Q\ldotp \lambda x\ldotp P~x\land Q~x$, which is exactly what is needed --- the grammar rules will apply this function to the semantics of the even and positive predicates, and finally 4.
Disjunction is handled similarly, and this generalizes to arbitrarily complex slash types whose final semantic result is \lstinline|Prop|.

\subsection{Using Specifications}
We couple these with an additional typeclass \lstinline|Semantics s C| which is defined 
when the string \lstinline|s| splits into a list of words \lstinline|w| (string splitting is implemented with another typeclass) and there exists an instance of \lstinline|Synth w C|.  Then we define a function from strings to their denotations:
\looseness=-1
\begin{lstlisting}[language=Coq]
Definition spec (s:string) `{sem:Semantics s S} := sdenote sem.
\end{lstlisting}
When invoked with a string \lstinline|s|, \coq will search for an instance of \lstinline|Semantics s S| --- a parse of the string as a complete sentence.  The logical form of a sentence has \coq type \lstinline|Prop| (a proposition, or logical claim, to be proven). 

Thus we may translate a range of specifications given an appropriate lexicon, including those below (sugared into math notation for space and readability):
\[%
\begin{array}{l}
\llbracket\textrm{addone is monotone}\rrbracket = \forall x,y : \mathbf{N}\ldotp x\le y\Rightarrow \mathsf{addone}\;x\le\mathsf{addone}\;y\\
\llbracket\textrm{every natural is non-negative}\rrbracket= \forall n:\mathbf{N}\ldotp n \ge 0\\
\llbracket\textrm{every natural is non-negative and some natural is even}\rrbracket= (\forall n:\mathbf{N}\ldotp n \ge 0)\land(\exists n:\mathbf{N}\ldotp \mathsf{even}\;n)\\
\end{array}
\]%
Because \coq's logic allows proof goals to be computed,
\lstinline|spec| can be used to declare proof goals:
\begin{lstlisting}[language=Coq]
Goal spec "addone is monotone".
> spec "addone is monotone"
simpl. (* simplify the goal *)
> forall x y : nat, x <= y -> addone x <= addone y
\end{lstlisting}
If \coq cannot find a \lstinline|Semantics| instance for a specification, the user sees an error; because instance search reuses existing proof search functionality, a skilled user could manually debug the failure.

  \subsection{Predicativity}
  A careful reader will notice that Section \ref{open-categorial-grammars-over-cic} defines a \textsf{Type} whose constructors quantify over \textsf{Type}.  Our current use cases produce elements of \textsf{Prop}, so we could simply define \lstinline|Cat| has having sort \textsf{Prop}.  However, there are many reasons to conduct proofs avoiding \textsf{Prop} even in specifications (e.g., homotopy type theory~\cite{bauer2017hott}).  Making all of the typeclass definitions, instances, and goals universe-polymorphic (\lstinline|Polymorphic|) allows us to remain predicative.

\subsection{Performance}
The performance of semantic parsing from natural language into formal specifications depends on both the underlying typeclass resolution procedure, as well as the space of derivations that must be explored during parsing.
Our tokenization code for splitting strings runs in linear time because there is at most one typeclass instance that can apply for each character of a string (i.e., depending on whether or not the character is a space).
So the rules which determine the search space are primarily the structural rules encoded in the \lstinline|Synth| typeclass instances, and the \lstinline|lexicon| instances.  Since most words have only one or a very small number of grammatical roles (in general, not just in our small prototype~\cite{hockenmaier2005ccgbank,hockenmaier2007ccgbank,hockenmaier2006creating}), lexicon ambiguity will not be a major driver of search costs. Instead, most costs come from exploring structural rule applications, particularly the rules for associativity and shifting of left and right slash type nesting, each of which may send search off into a dead end. To limit ambiguity, our typeclass for coordinators imposes an upper bound on how many times boolean operations can be lifted, and in Coq we set the typeclass resolution to use a depth limit of 15 on instance resolution. This reflects that by and large, categorial grammar derivations tend to be shallow and wide.

As a coarse measure of base efficiency, under these conditions parsing ``every natural is non-negative and some natural is even'' takes approximately 26 seconds\footnote{Measured by setting \lstinline|Semantics "every natural is non-negative and some natural is even" S| as the proof goal, and using \lstinline|time typeclasses eauto| to time the search.} on a 4-core Intel Core i5 (with 16GB of RAM, but this is CPU-bound). Faster would be ideal, but this is at least suitable for using one file to cache the parses, which can be compiled infrequently, with other files pulling in the relationships as needed. Inspection of the search trace confirms that most search time is spent in dead-ends related to different combinations of associativity and shifting rules.

To investigate whether alternative typeclass resolution algorithms might have an impact, we ported most of our machinery --- all except string tokenization\footnote{String tokenization takes negligible time in \coq for reasons outlined above, and does not directly translate because Lean uses a different datatype for strings. } --- to version 4 of the Lean theorem prover~\cite{moura2021lean}. This version uses a new typeclass resolution algorithm~\cite{selsam2020tabled} designed specifically to accelerate complex typeclass resolution problems like the one our work presents. Working with pre-tokenized inputs, compiling the entire formalization and parsing multiple examples, including the one above, consistently takes less than 3.5 seconds. %
More complex sentences of course may require more time and investment in optimization, as the simplest categorial grammars model the context-free languages (which have worst-case cubic time parsing), while the most complete mildly-context-sensitive classes of categorial grammars have $O(n^6)$ worst-case parsing cost~\cite{joshi:91}, though in practice the common case can be made quite fast~\cite{Clark:2007,Clark:2003,Clark:2002}.

\section{Modularity and Extension: Growing a Lexicon, Handling More Logics}\label{sec:treebanks}
The previous section described only a small fragment of English suitable for formalizing mathematical claims.
Categorial grammars are what linguistic semanticists call \emph{lexicalized} grammar formalisms. Unlike phrase-structure grammars (e.g., context-free grammars) which build in an explicit classification of grammatical phrase types, lexicalized grammars use a small set of general rules (like those in Figure \ref{fig:rules}), and then rely on the lexicon to give the precise grammatical types of every word. The availability of slash types (directed function types) affords significant flexibility, and extensions to attach modalities to the slashes~\cite{Baldridge:2003:MCC:1067807.1067836,Moortgat1996} allow further constraints capturing the subtleties of natural language to be captured solely by giving precise grammatical types (and semantics) to individual words.

\subsection{Managing Words}
Adding new words to a categorial grammar lexicon is conceptually as simple as adding the word, particular grammatical type, and associated denotation to the database.  This makes it easy to extend a system with new concepts (e.g., new algebraic structures); lexicon entries to deal with concepts defined in a proof assistant library can be distributed as a part of that library.
Conversely, if a word or particular usage of a word is found to be confusing to humans, leading to ambiguity, or otherwise problematic, it can be removed from the lexicon while affecting only inputs that use that word in that way (i.e., the problematic ones).

In practice the situation will be more complex, but we expect most extension to require little, if any, special linguistic knowledge. Assuming a robust core lexicon (Section \ref{sec:corelex}), it is likely that most extensions will be additions of words with simpler categories.
Experiments on a large standard-English lexicon showed~\cite{hockenmaier2005ccgbank} that when training on most of lexicon, the unseen words in a held-out test set were primarily nouns (35.1\%) or transformations of nouns (e.g., adjectives, at 29.1\%). These are the simplest categories to provide semantics for (types, objects, and predicates), strongly suggesting that proof assistant users with no special linguistics background could make most extensions themselves. Similar experiments for a wide-coverage lexicon of German~\cite{hockenmaier2006creating} show over half of unknown words to be nouns, suggesting this feasibility extends beyond just English.

\subsection{Supporting Additional Grammatical Constructions}
Formalization of significant fragments of language much deal with more subtle constructions that what we have described so far can handle. However, what we have described thus far is essentially read directly out of the literature on linguistic semantics.  
Linguists have spent many decades building out knowledge of how to handle more sophisticated uses of quantification~\cite{steedman2012taking,moortgat1996generalized} (``every,'' ``some,'' ``most''), resolving pronoun references~\cite{jacobson1999towards}, discontinuity~\cite{morrill1995discontinuity} (where a word is far from a word it modifies), and much more~\cite{carpenter1997type,morrill2012type}.

\subsection{A Full-Featured Core Lexicon}
\label{sec:corelex}
How large should a base lexicon with reasonably wide coverage be?
The largest lexicon for a categorial grammar is Hockenmaier and Steedman's CCGBank~\cite{hockenmaier2006creating,hockenmaier2007ccgbank}, which models the usage of language in a particular sample of the Wall Street Journal.  It contains roughly 75K unique words, and some of the most common words have dozens of grammatical categories, or more (the English word ``as'' is overloaded with 130 distinct --- though related --- grammatical types).  This has motivated work on learning lexicons with semantics~\cite{kanazawa1995learnable,artzi2013weakly,Zettlemoyer:2005,KwiatkowksiZGS10}, as well as work on learning more compact lexicons that automatically capture standard word variations (e.g., automatically generalizing singular definitions to work for plural forms)~\cite{LewisS14,wang2014morpho,Kwiatkowski2011}. These works all focus on cases where CCGs parse sentences into a variant of first-order logic, but should in principle generalize to targeting richer logics like the Calculus of Inductive Constructions underlying \coq and \textsc{Lean}.  While the need to scale to large lexicons draws us back to a kind of machine learning, it draws us back to a kind with an eminently auditable results, producing lexicon entries which have well-defined individual meaning, which can be manually adjusted or removed if necessary.

Any initial broad-coverage lexicon for technical prose will need to be manually constructed (including input to a future learning algorithm). However, since technical prose about math and code is still a particular stylistic use of a standard natural language, it mostly reuses words in the same grammatical role --- and therefore, same categorial grammar type --- as non-specialist grammars. This means we can bootstrap an initial lexicon for English by reusing grammatical types from existing categorial grammar lexicons for English~\cite{hockenmaier2007ccgbank,honnibal2010rebanking}, and similarly German~\cite{hockenmaier2006creating}, Hindi~\cite{ambati2018hindi}, Japanese~\cite{mineshima2016building}, and other languages~\cite{abzianidze2017parallel}.

This means initial efforts can focus mostly on defining semantics for existing grammar-only lexicon entries, rather than starting from scratch. And for many of the words appearing in specifications, particularly quantifiers (``every'', ``all'', etc.), determiners (``a'', ``the'', etc.), and prepositions (``in'', ``of'', etc.), the semantics are typically very simple (quantifier semantics look like the examples in earlier sections; prepositions are typically identity functions, functioning similarly to linguistic phantom types~\cite{fluet2006phantom} / units of measure~\cite{kennedy1994dimension} for other words to locate parameters.)

Careful readers or prior students of linguistics may have wondered when matters of verb tense, noun case and number, grammatical gender,\footnote{Which does not exist in English, but does in German, French, and so on} etc. would arise.  In full linguistic treatments, these are reflected in additional parameters to some grammatical categories.  So for example, in our setting a noun phrase would be parameterized not only by the underlying referent type, but also by the case, number and so on; lexicon entries would then carry these through appropriately (making it possible to for example, require the direct object of a verb to be in the accusative case rather than nominative). We have omitted such a treatment here partly because it would obscure the key ideas while adding little value, partly because many of these distinctions are less important for our examples in English (which has fewer syntactic case distinctions than other languages), and partly because some aspects (like tense) may make sense only for specific embedded specification logics.

\subsection{Beyond \textsc{CIC}}
\label{sec:generalization}
While the framing in this paper has focused on generating specifications which in \coq and Lean have type \lstinline|Prop|, this is not required. Categorial grammars require only that their top-level semantic truth value type have the structure of a Heyting Algebra~\cite{lambek1988categorial}: a type with binary operators for standard logical operators.

Our \coq and Lean formalizations in fact make this generalization: the core machinery is polymorphic over an arbitrary choice of Heyting Algebra, with a lexicon split between entries polymorphic over the Heyting Algebra being targeted (e.g., ``or'' and ``and'') and words specific to a given Heyting Algebra.

This means the core idea applies not only to specs of type Prop, but that this machinery can be readily retargeted to any logic formalized within the proof assistant, such as LTL~\cite{pnueli1977temporal}, CTL~\cite{clarke1986automatic}, or the BI-algebras underlying separation logics like Iris~\cite{jung2018iris}.

This is not itself a new observation, as we discuss in related work (Section \ref{sec:relwork}), as others have instantiated categorial grammars to generate, for example, CTL~\cite{dzifcak2009and}, before.  However, these prior applications have targeted only specific use cases, while this setting permits reusing many lexicon entries across \emph{many} logics, which should help in retargeting this machinery to new applications and future logics which may be formalized within a proof assistant.

  \section{Trust and Auditing}\label{trust-and-auditing}
  
  One of the essential criteria for an LCF-style proof assistant is the
  production of an independently-checkable proof certificate. While we
  have proposed using \coq's typeclass machinery to automatically parse and
  denote, and the typeclass resolution itself is typically not viewed as part of the trusted computing base (TCB), it does effectively produce a proof certificate.
  The typeclass machinery
  explicitly constructs a an instance of the typeclass --- an element of
  the corresponding record type --- and passes it to \lstinline|spec|.  So Coq's kernel sees (effectively) a categorial grammar proof, constructed via typeclass instances rather than constructors of an inductive data type.
   This explicit
  term persists into the proof certificates Coq already produces, and
  could be identified by an independent proof checker that wished to also
  validate the natural language interpretation.
  For example, the textual representation of the term witnessing that ``four is even'' parses to \lstinline|even 4| is:
\begin{lstlisting}
bridgeStringWords (* The typeclass instance to tokenize then parse *)
  (split1 NotSpace4 (* Start of tokenization *)
          (split2 NotSpace6 NotSpace6 
                  (split4' NotSpace6 NotSpace6 NotSpace6 NotSpace6)))
  (SynthLApp (SynthLex fourlex) (* Parse tree of tokenized string *)
     (SynthRApp (SynthShift (SynthLex noun_is_adj_sentence))
                (SynthLex even_lex)))
\end{lstlisting}
which encodes both the tokenization (\lstinline|split1|, etc.) and the constructed parse tree (with \lstinline|SynthLex| calls referring to individual lexicon instances).

  We can think of several ways a user might accidentally or maliciously risk confusing
  an independent checker:
  All but one can easily be detected by a checker aware of the categorial grammar specification typeclasses.
  The final possibility amounts to changing the specification in the proof certificate.
  
  \begin{itemize}
  \item
    A user may redefine or extend our core instances (for \texttt{Synth}) to produce a different denotation.
    A certificate checker would already ensure these are type-correct. An natural-language-specification-aware extension could check that the \lstinline|Synth| instances correspond to the desired rules.
    Or to better support some of the extensibility arguments made earlier, the \lstinline|Synth| typeclass could be modified to also carry a justification of its conclusions in a more general substructural logic~\cite{kruijff2000relating,gerhard2005anaphora}, which would amount to requiring extensions to carry conservativity proofs.
  \item
    A user may extend the lexicon with additional words or additional grammatical roles for a given word, introducing ambiguity into the parsing.  Checking for ambiguity is relatively straightforward: ignoring indexing by \coq types, equivalence of grammatical types is decidable, and a checker could conservatively require that any lexicon entries with the same index-erased grammatical types have clearly-distinct indices.  An independent checker could verify the absence of ambiguity in the lexicon, or alternatively surface the use of any ambiguity in a parsing derivation for human inspection.
  \item
    A user could also manipulate the lexicon, for example \emph{redefining} ``monotone''
    to denote as \(\lambda f, \mathsf{True}\). This is arguably a form of
    modifying the specification by changing definitions, rather than
    sneaking a broken proof past a certificate checker.  It is analagous to changing a definition of a property verified by a proof --- a working proof with the wrong definition is wrong, but this leaves behind evidence of the incorrect definition.
  \end{itemize}

These possible forms of attack highlight the main sources of trust added when considering natural language specifications in the approach we describe: the grammatical rules for combining phrases, well-formedness of the lexicon, and the definitions of words in the lexicon.

A minor point of trust is the requirement that lexicon entries at least give semantics whose \coq type is consistent with the grammatical type at hand. In our prototype this is expressed via the requirement that the type of a word's denotation is given by applying the \lstinline|interp| function to the grammatical type. This is checked automatically by encoding this requirement in \coq's types, and enforced by any proof certificate checker for \coq's logic. This might seem trivial, but it is worth noting because other implementations of categorial grammars often do not check types.  In early experiments we encountered type-related bugs in NLTK's \textsc{CCG} implementation, and our attempt to directly reproduce an existing use of \textsc{CCG}s for temporal logic specifications~\cite{dzifcak2009and} failed not because of our prototype's limitations, but because we encountered cases where the published lexicon entries were inconsistent with the stated grammatical types (e.g., giving a word a grammatical type indicating two arguments, but a logical form which only accepted one). It is possible these were merely typographical errors, as is often found by any mechanization based on a published paper rather than a code artifact~\cite{klein2012run}, but in either event the type system detected the incorrect entries when we attempted to enter them directly from the paper.

\section{A Limited Role for Machine Learning}
As discussed earlier, the need for predictability, auditability, and modular lexicons play to the strengths of categorial grammar. These also correspond to established weaknesses of common statistical and neural approaches to natural language processing, which often handle equivalent expressions in surprisingly incompatible ways (despite intriguing ongoing work to alleviate this~\cite{zhang-etal-2021-certified}), provide no interpretable justification for linguistic choices, and are inherently non-modular (one cannot simply retrofit a small collection of additional words into a large model).
Beyond this, neural approaches appear to systematically struggle to deal consistently and accurately with boolean operations (especially negation)~\cite{traylor2021and,ettinger2020bert,pandia-ettinger-2021-sorting}, which are often critical to formal specifications.

One of the primary advantages of neural and statistical models for natural language processing is that natural language is constantly evolving, with new expressions and terms being invented regularly, and neural and statistical models can often handle unseen words somewhat reasonably by mimicking known usage patterns of other words, despite lacking any ground notion of meaning~\cite{bender2020climbing,merrill2021provable}. However, this advantage has little role to play for formal specifications, both because of the hightened certainty requirements, but because new terms often carry very specific meanings, and also because the natural language used to describe formal properties tends to evolve more conservatively in order to avoid human misunderstandings.

The biggest general advantage of machine learning techniques is their ability to process large amounts of data with reduced human effort. Historically, large databanks of grammatical types and semantics for English~\cite{hockenmaier2007ccgbank}, German~\cite{hockenmaier2006creating}, Hindi~\cite{ambati2018hindi}, Japanese~\cite{mineshima2016building}, French~\cite{moot2015type}, and other languages~\cite{abzianidze2017parallel} have been rooted in enormous human efforts to manually label data, either directly or via translation from a corpus manually annotated for another grammar formalism.
This has inspired a range of techniques for learning a lexicon for semantic parsing from a small set of initial examples~\cite{Zettlemoyer:2005,artzi2013weakly,KwiatkowksiZGS10,Kwiatkowski2011}. These techniques could in principle be used to rapidly expand an initially-hand-crafted core lexicon, or eventually to learn a domain- or program-specific lexicon to be added to a base lexicon after auditing.
Currently all of these techniques target learning logical forms in first-order logic, and would require adaptation to deal with the indexed grammars required to target a proof assistant's logic.

One way machine learning could play a major role in this endeavour would be analagous to one of its major roles in traditional semantic parsing, which is in learning an optimal search strategy over derivations from a large corpus of complete derivations~\cite{Clark:2002,Clark:2003,Clark:2007}. Such a statistical model over likely parse structures can be used to dramatically accelerate parsing by using a model to choose priorities for proof search, thereby avoiding more dead-ends. Using machine learning in this narrow way could lead to substantial performance gains without compromising the categorial grammar properties relevant for formal specifications: a successful parse still yields a full derivation.
In principle it should be possible to learn how to assign rule priorities and/or customize generate custom (derived) rules.
However, a prerequisite for training such models is a large corpus of complete parses, which at least initially would need to be obtained through regular unification alone.

\section{Related Work}\label{related-work}
\label{sec:relwork}
Both categorial grammars of the form we work with and the use of dependent type theories for natural language semantics have long histories~\cite{lambek1958mathematics,vanBenthem90,sundholm1986proof}, and we are hardly the first to propose reducing the gap between formal or semi-formal specifications and natural language. Our proposal differs from the former primarily in that we argue for employing these not for the study of linguistics, but for the application of linguistics research to build a system for integrating natural language descriptions into the main intended use of proof assistants, including cases where the proof assistant is used to construct proofs in an embedded logic. Our proposal differs from most work on the latter primarily in our focus on employing actual linguistic models of grammar and meaning to extract intent from natural language, rather than using a range of shallow (though often effective) heuristics, in order to afford a higher degree of freedom in expressing expectations.
We offer more details on these relationships below.

Others have used type theories like \coq's for logical forms~\cite{sundholm1986proof,ranta1991intuitionistic,ranta1994ttg,chatz2014nlcoq,bekki14dts}, broadly making the argument that variants of dependent type theory offer a range of appealing options for modeling natural language semantics that fix some percieved deficiencies in the use of a lambda calculus over first-order logic formulas, but consistently focused on using this as a means to study linguistics.
The notion of indexing some grammatical categories by the type of a referent in such an underlying type theory comes from Ranta's work~\cite{ranta1995context} on studying the linguistics of mathematical statements.
Ranta~\cite{ranta1994ttg}, Kokke~\cite{kokke2015agda} and Kiselyov~\cite{kiselyov2015applicative} have formalized variants of categorial grammar with semantics in proof assistants, but only as object logics of study in order to prove properties of those systems, rather than as working parsers integrated with other uses of proof assistants. %
This leaves much to explore in integrating categorial grammar with various forms of type-theoretical language semantics~\cite{chatzikyriakidis2017modern}, some of which coincide with common specification patterns.

Others have worked towards using categorial grammars and related techniques to translate natural language into formal specifications in a variety of other logics. Dzifcak et al.~\cite{dzifcak2009and} used \textsc{CCG} to translate natural language specifications to $\textrm{CTL}^*$, though as mentioned in Section \ref{sec:generalization} their semantics contain semantic type errors which are caught by working within a proof assistant that \emph{enforces} consistency between grammatical and semantic types.
Seki et al.~\cite{seki1988processing,seki1992method} is the earliest approach we are aware of, using an alternative grammar formalism (HPSG~\cite{pollard1994head}) to translate natural langauge to first-order logic.
Each of these approaches targets only a single logic, and assumes the translation is divorced from any particular use of formal specifications.

There have also been notable attempts to translate formal specifications into English, such as the support in the KeY theorem prover~\cite{johannisson2007natural}, and a predecessor system~\cite{burke2005translating} that directly uses categorial grammar in reverse for natural language generation~\cite{ranta2004grammatical} (another established use for categorial grammars beyond the semantic parsing we focus on).

There are also other approaches to bringing rigorous formalization closer to natural language, without attempting to capture natural language grammar in a systematic way.
Isabelle/HOL's~\cite{nipkow2002isabelle} Isar proof language~\cite{wenzel1999isar} attempts to make proofs themselves more readable using proof manipulation commands resembling English. This is an example of what is known as \emph{controlled natural language}~\cite{fuchs2009controlled}, a pattern of system development where the input language is a heavily restricted fragment of natural language, usually (though not always) simple enough to enable fully automatic processing, usually by heavily restricting both grammatical constructions and vocabulary.
This includes examples like Cramer et al.~\cite{cramer2009naproche}, who have worked on heavily restricted subsets of natural language that address both the specification of lemmas and their proofs, but like Isar do not attempt to capture any general natural language structure, and techniques which focus only on stating formulas in natural language~\cite{fuchs1999controlled}.
By contrast, our proposed approach (1) reuses existing proof assistant machinery (typeclasses~\cite{sozeau2008first,selsam2020tabled}) rather than requiring specialized support, and (2) aims to (eventually) permit almost arbitrary natural language grammar once an adequate base lexicon is developed (which can then be directly extended by individual proof developments).

\section{Looking Forward}
We have presented evidence that it is plausible to support natural language specifications in current proof assistants by exploiting existing typeclass machinery, with no additional tooling required.  Carried further, this could be useful in many ways.  It can reduce the gap between informal and formal specifications, reducing (though not eliminating) trust in the manual formalization of requirements.  Potentially non-experts in verification could understand some theorem statements, gaining confidence that a verification result matched their understanding of desired properties. And this could be used in educational contexts to help students learn or check informal-to-formal translations.

Of course, the details matter as well, and it will take time to realize a prototype that is broadly useful. First and foremost, a rich lexicon is required. As explained earlier, at least the initial lexicon will need to be manually constructed (borrowing grammatical categories from existing lexicons, and filling in the semantics) before it would be fruitful to adapt techniques for learning lexicons. Guiding this effort would require a substantial collection of examples of natural-language descriptions of formal claims, both for prioritizing lexicon growth and for validation that the approach is growing to encompass real direct descriptions of claims. 
Despite the now-enormous body of formalized proofs of program properties and mathematical results, 
early efforts in this direction have revealed this is less trivial than it seems. Even popular and classic texts introducing formal specification like Software Foundations~\cite{sf} and classic texts like Type Theory and Functional Programming~\cite{ttfp} have remarkably few crisp natural language statements matching a specific formal statement, instead discussing various needs at length in order to motivate eventual details of the final formalization (which is sensible for expository texts). Reynolds' classic paper introducing separation logic~\cite{reynolds02} contains no English-language description of any full invariant involving separating conjunction.
We do not necessarily require exemplars of full descriptions in a single sentence; it is common for one specification to imply multiple high-level properties, and we envision one style of use for natural language specifications to be checking that a given verified result implies multiple natural language claims which each cover part of the desired results.

It is possible that small differences will be required between standard natural language grammars and those used by this approach, arising from distinctions important to proof assistants but irrelevant to colloquial language. This is already the case, as mentioned, with the indexing of some grammatical categories with the semantic types of referents, following Ranta's early work on formalizing mathematical prose~\cite{ranta1995context}.
This direction offers opportunities to collaborate with linguists working in syntax and compositional semantics~\cite{barker2007direct,jacobson2014compositional,steedman2012taking}. Such collaborations could both help with possible novel linguistic features of ``semi-formal'' natural language, and offers a setting for applying classical linguistic techniques in a domain where they provide unique value.

A great deal of work lies ahead, but the potential benefits seem to more than justify further exploration in this direction.

\end{document}